\documentclass[sigconf]{acmart}
\usepackage{multirow}
\usepackage{tabularx}
\usepackage{float}

\copyrightyear{2026}
\acmYear{2026}
\setcopyright{cc}
\setcctype{by}
\acmConference[WWW '26]{Proceedings of the ACM Web Conference 2026}{April 13--17, 2026}{Dubai, United Arab Emirates}
\acmBooktitle{Proceedings of the ACM Web Conference 2026 (WWW '26), April 13--17, 2026, Dubai, United Arab Emirates}
\acmPrice{}
\acmDOI{10.1145/3774904.3792174}
\acmISBN{979-8-4007-2307-0/2026/04}

\begin{document}

\title{Explaining Synergistic Effects in Social Recommendations}


\author{Yicong Li}
\authornote{Both authors contributed equally to this research.}
\affiliation{%
  \institution{Dalian University of Technology}
  \city{Dalian}
  \country{China}}
\email{ycongli@outlook.com}

\author{Shan Jin}
\authornotemark[1]
\affiliation{%
  \institution{Dalian University of Technology}
  \city{Dalian}
  \country{China}}
\email{jinshan0924@mail.dlut.edu.cn}

\author{Qi Liu}
\affiliation{%
  \institution{Dalian University of Technology}
  \city{Dalian}
  \country{China}}
  \email{1462255303@mail.dlut.edu.cn}

\author{Shuo Wang}
\affiliation{%
  \institution{Dalian University of Technology}
  \city{Dalian}
  \country{China}}
\email{ws123416@mail.dlut.edu.cn}

\author{Jiaying	Liu}
\affiliation{%
  \institution{Dalian University of Technology}
  \city{Dalian}
  \country{China}}
\email{jiayingliu@dlut.edu.cn}

\author{Shuo Yu}
\authornote{Corresponding author.}
\affiliation{%
  \institution{Dalian University of Technology}
  \city{Dalian}
  \country{China}}
  \email{shuo.yu@ieee.org}

\author{Qiang Zhang}
\authornotemark[2]
\affiliation{%
  \institution{Dalian University of Technology}
  \city{Dalian}
  \country{China}}
\email{zhangq@dlut.edu.cn}

\author{Kuanjiu	Zhou}
\affiliation{%
  \institution{Dalian University of Technology}
  \city{Dalian}
  \country{China}}
\email{zhoukj@dlut.edu.cn}

\author{Feng Xia}
\affiliation{%
  \institution{RMIT University}
  \city{Melbourne}
  \country{Australia}}
\email{f.xia@ieee.org}

\renewcommand{\shortauthors}{Yicong Li et al.}

\begin{abstract}
In social recommenders, the inherent nonlinearity and opacity of synergistic effects across multiple social networks hinders users from understanding how diverse information is leveraged for recommendations, consequently diminishing explainability. 
However, existing explainers can only identify the topological information in social networks that significantly influences recommendations, failing to further explain the synergistic effects among this information.
Inspired by existing findings that synergistic effects enhance mutual information between inputs and predictions to generate information gain, we extend this discovery to graph data. We quantify graph information gain to identify subgraphs embodying synergistic effects.
Based on the theoretical insights, we propose SemExplainer, which explains synergistic effects by identifying subgraphs that embody them.
SemExplainer first extracts explanatory subgraphs from multi-view social networks to generate preliminary importance explanations for recommendations.
A conditional entropy optimization strategy to maximize information gain is developed, thereby further identifying subgraphs that embody synergistic effects from explanatory subgraphs.
Finally, SemExplainer searches for paths from users to recommended items within the synergistic subgraphs to generate explanations for the recommendations.
Extensive experiments on three datasets demonstrate the superiority of SemExplainer over baseline methods, providing superior explanations of synergistic effects\footnote{The implementation is available at {\url{https://github.com/yushuowiki/SemExplainer}}}.

\end{abstract}

\ccsdesc[500]{Information systems~Social recommendation}
\ccsdesc[500]{Human-centered computing~Social networks}
\keywords{Social Networks, Social Recommendations, Synergistic Effects, Explainability, Multi-view Graphs.}


\maketitle

\section{Introduction}
\begin{figure}[t]
  \centering
  \includegraphics[width=\linewidth]{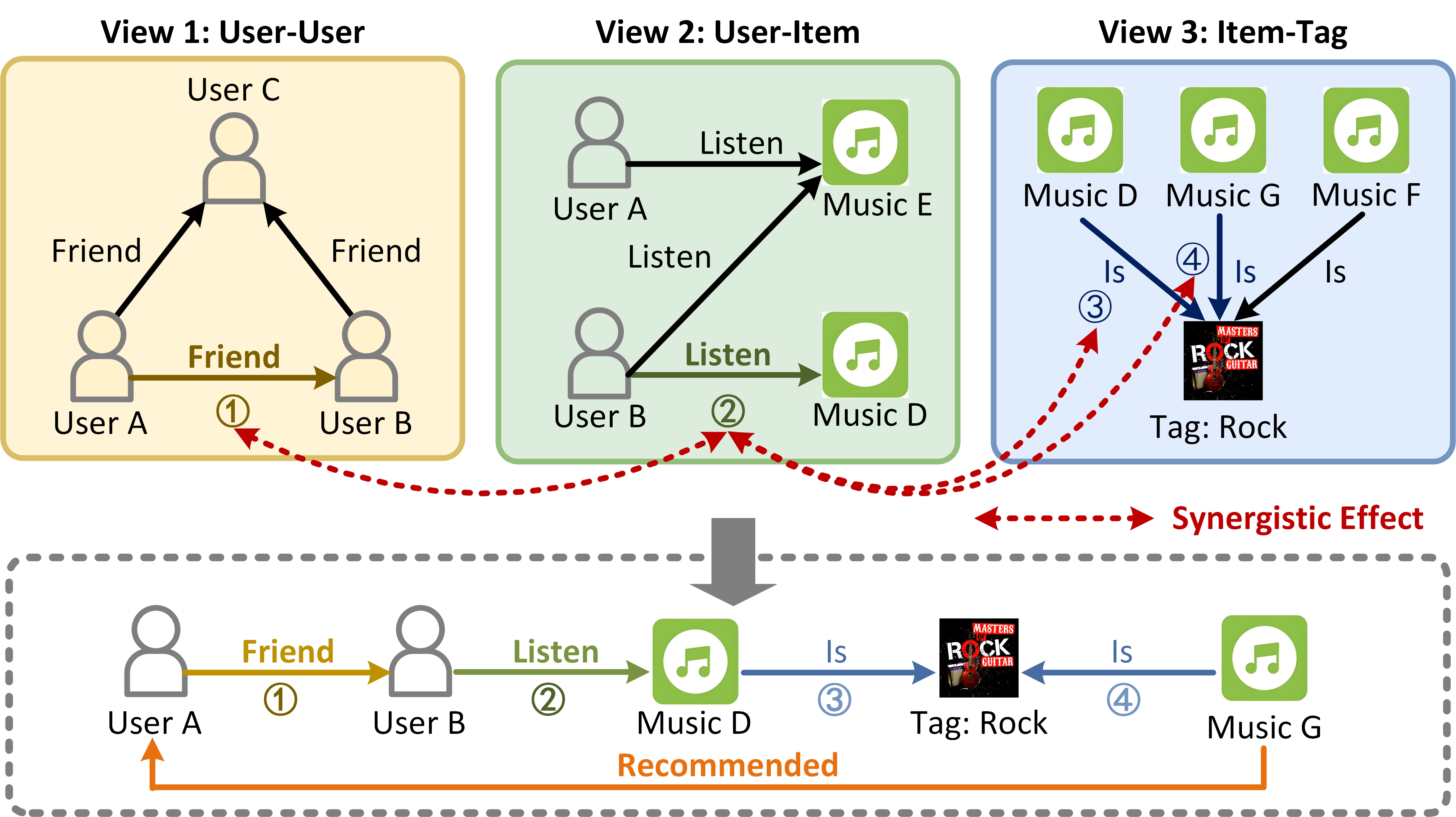}
  \caption{The illustration of synergistic effects. The system synergistically integrates information from three views: (1) User A and User B are friends; (2) User B listens to Music D; and (3) Music D and Music G both belong to the rock genre. Through the interaction of these views, the system infers that User A is likely to enjoy Music G.}
  \vspace{-2em}
  \Description{An example of synergistic effects in music recommendation.}
  \label{fig:intrduction}
\end{figure}

Social recommendations have substantially improved the dissemination and distribution of information across the World Wide Web.
Synergistic effects arising from multi-view information can substantially enhance the performance of social recommendations, particularly on social platforms that involve analyzing multiple types of relations between users and items~\cite{kim2024discovering, xiong2025robust, zheng2024inductive}.
Figure~\ref{fig:intrduction} illustrates the synergistic effects in social recommendation tasks.
The social recommendation system synergistically analyzes interactions among social relationships across multiple views, uncovering valuable insights that cannot be captured from a single view, thereby producing more accurate recommendations~\cite{hao2025gnnsynergy, 10814649, yang2025towards}.
%
However, the inherent nonlinearity and opacity of synergistic effects make it difficult to explicitly explain these recommendation results~\cite{10597965, 10.1145/3626243, chen2025semi}.
This naturally raises concerns regarding the reliability of social recommendation systems~\cite{10.1145/3626772.3657967, 10742303, chen2024deep}.
The inability to understand synergistic effects may lead to user confusion about recommendations~\cite{lifactor}.
The black-box nature of synergistic effects also prevents developers from assessing the contributions of different views, thereby limiting the optimization of system architecture.
Therefore, it is crucial to improve the explainability of social recommendation systems through an explicit explanation of synergistic effects.

Numerous efforts have been proposed to explain the results for social recommendation systems.
Their mainstream paradigm focuses on extracting key feature subsets, such as subgraphs~\cite{chen2024feature, 10.1145/3589334.3645608, 10.1145/3589334.3645599} or paths~\cite{10463173, 10399934, yucage} within multi-view topological information networks.
Empirical results show that these methods can accurately and succinctly convey the rationale behind recommendation decisions to users~\cite{guan2025explainable, medda2024gnnuers, zheng2025dp}, but fail to explain the synergistic effect between different views.
The underlying reason is that information influencing recommendation results does not necessarily interact synergistically with other information. Extracting a subset of key features fails to reveal which features are used synergistically~\cite{nian2024globally, akkas2024gnnshap, sinha2025higher, dygmf}.
Moreover, synergistic effects emerge during the implicit reasoning process of recommendation, making them technically challenging to observe and explain explicitly~\cite{guan2025structure, dong2025enhanced,luo2024fairgt}.
Inspired by existing finding~\cite{kim2024discovering} that synergistic effects enhance mutual information between inputs and predictions to generate information gain, we extend this finding to graph data and propose the synergistic subgraph theory. 
The objective is to identify subgraphs embodying synergistic effects by quantifying the information gain in graph mutual information~\cite{wei2022contrastive, yu2024long}.
To this end, we first define the concept of synergistic subgraphs, which formalizes the mutual information conditions that must be satisfied by topological information embodying synergistic effects.
Next, we formalize a single inequality constraint for the information gain of synergistic subgraphs based on mutual information theory and formulate a conditional entropy optimization problem to enable the extraction and explanation of synergistic subgraphs.

On this basis, we propose SemExplainer, a synergistic effect explainer for social recommendation systems.
SemExplainer consists of three components, including explanatory subgraph extraction, synergistic information disentanglement, and path explanation generation.
In the first module, SemExplainer utilizes mask learning to extract key subgraphs that significantly influence recommendation results from each view graph, generating preliminary explanatory subgraphs for multi-view graph neural networks~(GNNs).
Afterwards, explanatory subgraphs are partitioned into synergistic and non-synergistic subgraphs. 
A conditional entropy optimization strategy is designed to maximize the information gain of synergistic subgraphs while minimizing that of non-synergistic subgraphs.
This ensures the disentanglement of synergistic information within the explanatory subgraphs. Finally, SemExplainer searches for paths from users to target items on multi-view graphs, constraining the paths to traverse nodes and edges that contain high synergistic information.
These paths provide concise and readable explanations for the synergistic effects of multi-view GNNs in social recommendation.
Our contributions are outlined as follows.

\begin{itemize}
\item {\textbf{Theory for Discovering Synergistic Effects:}}~we propose the synergistic subgraphs theory, which provides a theoretical basis for judging whether topological information produces synergistic effects.

\item {\textbf{Explanation of synergistic effect:}}~we present SemExplainer, an innovative multi-view GNN explainer that can explain synergistic effects between different types of topological information in social recommendation tasks. 
It reveals core reasons for multi-view GNN boosting recommendation performance, instead of merely extracting feature subsets that influence predictions.
\item {\textbf{Strategy to distinguish synergistic effect:}}~we devise a mask learning method based on conditional entropy optimization to extract synergistic subgraphs by maximizing information gain.
Unlike previous methods that merely generate explanatory subgraphs, SemExplainer further extracts information that embody synergistic effects from explanatory subgraphs.
\item {\textbf{Effectiveness in explaining synergistic effects:}}~Extensive experimental results show that SemExplainer is more effective in explaining synergistic effects in social recommendation tasks than SOTA baselines. SemExplainer can reveal the underlying rationale behind how social recommenders synergistically use different types of social networks to make recommendations.
\end{itemize}

\section{Related Work}
\subsection{Explainable Social Recommenders}
Explainable social recommenders can be divided into two categories: GNN-generic explainers and system-specific explainers.
Since social recommendation models typically employ GNNs to encode social networks, existing GNN-generic explainers can be applied to explain social recommendation systems.
GNN-generic explainers aim to identify substructures of the input topology that exert the greatest influence on GNN predictions. 
These extracted substructures can, in turn, reveal the key factors driving the outcomes of social recommendation models.
For example, GNNExplainer~\cite{ying2019gnnexplainer} adopts masking optimization to filter subgraphs from individual instances, while PGExplainer~\cite{luo2020parameterized} uses generative models to produce subgraphs. Yu et al.~\cite{yumage} further investigate the use of motifs as fundamental units.
Beyond studies on explanation forms, Muschalik et al.~\cite{muschalikexact} introduce the use of the Shapley value to quantify the contributions of interactions among multiple nodes to the GNN predictions.
Liu et al.~\cite{liu2025explanations} investigate how variations in edge weights within graphs influence the GNN predictions.
The aforementioned methods can be readily extended to multi-view GNNs by iteratively optimizing each view.

As a complement to GNN explainers, system-specific explainers focus on explaining recommendation tasks. 
They frame recommendation behavior as a link prediction problem and explain why a particular item is recommended to a social user.
Zhang et al.~\cite{zhang2023page} utilize paths to explain the link prediction between users and items.
To uncover the causality between topological information and recommendation outcomes, Yu et al.~\cite{yucage} employ backdoor adjustment techniques to generate causal path explanations.
xPath~\cite{li2023towards} uses the greedy search strategy to generate path explanations, revealing the underlying relationships between user nodes and item nodes.
Furthermore, to allow explanations to inversely optimize recommendation performance, Huang et al.~\cite{huang2025social} propose graph-level counterfactual explanations to enhance social recommendation systems.
However, the aforementioned explainers assume that the contributions of topological information from different views to social recommendation are independent, which prevents them from explaining the synergistic effects across multiple views in social recommendation systems.

\subsection{Synergistic Effects of Multi-view GNNs}
Synergistic effects are one of the fundamental factors driving the superior performance of multi-view GNNs. 
Studies on synergistic effects in multi-view GNNs can be broadly categorized into two aspects: theoretical studies and application studies.
From the theoretical aspect, Kim et al.~\cite{kim2024discovering} elucidate the necessity of synergistic effects for enhancing the performance of multi-view learning.
Zhuang et al.~\cite{10.1145/3664647.3681585} propose a multi-view graph message-passing mechanism to strengthen the synergistic interactions across graphs from different views.
Jiao et al.~\cite{jiao2025deep} propose a self-enhancing view weight method that facilitates the synergy and fusion of multiple view graphs by learning the importance of each view.
From the application aspect, by leveraging synergistic effects, multi-view GNNs have achieved remarkable success in social recommendation scenarios.
For example, Luo et al.~\cite{10.1145/3649436} use multi-view GNNs to capture synergistic information across different user behavior sequences.
Ma et al.~\cite{ma2024multicbr} employ multi-view graph contrastive learning to identify synergistic interactions between users and items, thus improving the performance of the recommendation.
Xiong et al.~\cite{xiong2025robust} utilize multi-view GNNs to extract synergistic interaction patterns across various types of social information among users.
Wang et al.~\cite{10.1145/3726302.3729978} model both the periodic variations and the synergistic patterns across different types of user behavior.
Yang et al.~\cite{DBLP:conf/sigir/YangWLHSHW25} propose social graph invariant learning to identify stable synergistic interaction patterns across multiple views.
Despite the crucial role of synergistic effects in improving recommendation performance, their internal processes remain opaque, making it difficult to provide explicit explanations to users. Consequently, explaining synergistic effects remains a central challenge in achieving explainable social recommendations.

\section{Preliminaries}\label{sec:Preliminaries}
In social recommendation tasks, the recommendation predictions of the multi-view GNN model (with $h$ layers) rely exclusively on the $h$-hop neighborhood computational graph of user node $u$ and item node $i$. We correspondingly define the neighborhood computational graph for the $k$-th view as a directed graph ${\mathcal G}^{k} = ( {\mathcal V,\mathcal E^{k}} )$, where $\mathcal V$ denotes the set of nodes and $u,i \in \mathcal V$. 
$\mathcal E^{k}$ denotes the set of edges for the $k$-th view, which describes a type of social behaviors between nodes.
Let $\mathbf{X}\in {\mathcal{R}^{n \times d}}$ denote the node feature matrix, where $n$ is the number of nodes and $d$ is the feature dimension.
$\mathbf{X}$ contains the attributes of both users and items.
Let $\mathbf{A}^{k}\in {\mathcal{R}^{n \times n}}$ denote the adjacency matrix from the $k$-th view.
Let ${\mathcal G}=\{{\mathcal G}^{1},...,{\mathcal G}^{K}\}$ denote the set of graphs from $K$ views.
We use ${\mathcal G}^{\backslash k}=\{{\mathcal G}^{1},...,{\mathcal G}^{k-1},{\mathcal G}^{k+1},...,{\mathcal G}^{K}\}$ to represent the graphs from all other views except the $k$-th view.
Let $y$ denotes the recommendation prediction.

\paragraph{Definition~3.1 Synergistic Effect.}
In a social recommendation scenario, suppose that a user performs two social behaviors $e_1$ and $e_2$, leading to a recommendation prediction $y$.
The synergistic effect between $e_1$ and $e_2$ is quantified in terms of the difference between $I(y;e_1\left| e_2 \right.)$ and $I(y;e_1)$, where $I(y;e_1)$ denotes the mutual information between $y$ and $e_1$, and $I(y;e_1\left| e_2 \right.)$ denotes the conditional mutual information between $y$ and $e_1$ given $e_2$~\cite{kim2024discovering, dong2024multi, xu2024reliable}.
Specifically, we define $e_1$ and $e_2$ to have a synergistic effect if and only if they satisfy the following inequality:
\begin{equation}
\label{inequality}
I(y;e_1\left| e_2 \right.) > I(y;e_1)
,
\end{equation}
if inequality~(\ref{inequality}) does not hold, there is no synergistic effect between $e_1$ and $e_2$.
According to Definition 3.1, $e_1$ and $e_2$ have a synergistic effect if making $e_2$ known increases the mutual information between $e_1$ and $y$.

\section{Theoretical Discovery}
\subsection{Synergistic Subgraphs}\label{section4.1}
In multi-view graphs, we define the synergistic information of a given view as a subgraph of its input graph.
We define $\mathcal{G}^k_S$ as the synergistic subgraph of the $k$-th view.
The adjacency matrix of $\mathcal{G}^k_S$ is denoted as $\mathbf{A}^k_s = \mathbf{A}^k \odot \mathbf{M}^k$, where $\mathbf{M}^k \in[0,1]^{n\times n}$ is the edge mask matrix.
Similar to Section~\ref{sec:Preliminaries}, we define ${\mathcal G}_S$ as the set of synergistic subgraphs for all views, and define ${\mathcal G}^{\backslash k}_S$ as the set of synergistic subgraphs except for the $k$-th view.
To extract the synergistic subgraph, we extend Definition~3.1 to the multi-view graph setting. 
We define the synergistic subgraph of the $k$-th view as one that satisfies the following inequality:
\begin{equation}
I(y;{\mathcal G}^{k}_S\left| {{\mathcal G}^{\backslash k}_S} \right.) > I(y;{\mathcal G}^{k}_S)
.
\label{eq:graph-inequality}
\end{equation}
In other words, for each view, we aim to search for ${\mathcal G}^{k}_S$ such that inequality~(\ref{eq:graph-inequality}) holds.
This means that we need to optimize $K$ views at once through $K$ inequality constraints. 
To simplify the optimization, we utilize a single constraint to replace the multiple constraint problems defined for each view above to search for the synergistic subgraph set ${\mathcal G}_S=\{{\mathcal G}^{1}_S,...,{\mathcal G}^{K}_S\}$ for all views.
To this end, we introduce the following lemma.
\paragraph{Lemma 1.}
Assuming that ${\mathcal G}_S$ satisfies the following inequality:
\begin{equation}
I(y;{\mathcal G}_S) > \sum\nolimits_{k = 1}^K {I(y;{\mathcal G}^{k}_S)} 
,
\end{equation}
then any ${\mathcal G}^{k}_S \in {\mathcal G}_S$ also satisfies Inequality~\ref{eq:graph-inequality}.
\paragraph{Proof.}
We utilize the chain rule~\cite{wei2022contrastive} of graph mutual information to derive the following equivalent relationship.
\[
\begin{aligned}
&I(y;\mathcal{G}^k_S | \mathcal{G}^{\backslash k}_S) > I(y;\mathcal{G}^k_S) \\
\Leftrightarrow &H(y | \mathcal{G}^{\backslash k}_S) - H(y | \mathcal{G}^k_S, \mathcal{G}^{\backslash k}_S) > H(y) - H(y | \mathcal{G}^k_S) \\
\Leftrightarrow &- H(y | \mathcal{G}^k_S, \mathcal{G}^{\backslash k}_S) > H(y) - H(y | \mathcal{G}^k_S) - H(y | \mathcal{G}^{\backslash k}_S) \\
\Leftrightarrow &H(y) - H(y | \mathcal{G}^k_S, \mathcal{G}^{\backslash k}_S) > 2H(y) - H(y | \mathcal{G}^k_S) - H(y | \mathcal{G}^{\backslash k}_S) \\
\Leftrightarrow &I(y;\mathcal{G}^k_S, \mathcal{G}^{\backslash k}_S) > I(y;\mathcal{G}^k_S) + I(y;\mathcal{G}^{\backslash k}_S) \\
\Leftrightarrow &I(y;\mathcal{G}_S) > I(y;\mathcal{G}^k_S) + I(y;\mathcal{G}^{\backslash k}_S)
,
\end{aligned}
\]
where $H(\cdot)$ denotes entropy.
Furthermore, we discuss $I(y;\mathcal{G}^k_S) + I(y;\mathcal{G}^{\backslash k}_S)$ in two cases.
First, if ${\mathcal G}^{1}_S,...,{\mathcal G}^{K}_S$ are mutually independent, the following derivation holds:

\[I(y;\mathcal{G}^{\backslash k}_S) = I( {y;{ \cup _{j \ne k}}I(y;\mathcal{G}^j_S)} ) = \sum\nolimits_{j \ne k} I(y;\mathcal{G}^j_S). \]
Therefore, we obtain:
\[
\begin{aligned}
I(y;\mathcal{G}^k_S) + I(y;\mathcal{G}^{\backslash k}_S)& = I(y;\mathcal{G}^k_S) + \sum\nolimits_{j \ne k} I(y;\mathcal{G}^j_S) \\
& = \sum\nolimits_{k = 1}^K {I(y;{\mathcal G}^{k}_S)} 
.
\end{aligned}
\]
Second, if ${\mathcal G}^{1}_S,...,{\mathcal G}^{K}_S$ are not mutually independent, the following derivation holds:
\[I(y;\mathcal{G}^{\backslash k}_S) \le  \sum\nolimits_{j \ne k} I(y;\mathcal{G}^j_S). \]
Therefore, we obtain:
\[
\begin{aligned}
I(y;\mathcal{G}^k_S) + I(y;\mathcal{G}^{\backslash k}_S)& \le I(y;\mathcal{G}^k_S) + \sum\nolimits_{j \ne k} I(y;\mathcal{G}^j_S) \\
& \le \sum\nolimits_{k = 1}^K {I(y;{\mathcal G}^{k}_S)} 
.
\end{aligned}
\]
Combining the above two cases, we obtain:
\begin{equation}
I(y;{\mathcal G}^{k}_S) + I(y;{\mathcal G}^{\backslash k}_S) \le \sum\nolimits_{k = 1}^K {I(y;{\mathcal G}^{k}_S)} 
.
\end{equation}
Therefore, if the following inequality holds:
\begin{equation}
I(y;{\mathcal G}_S) > \sum\nolimits_{k = 1}^K {I(y;{\mathcal G}^{k}_S)},
\end{equation}
then $I(y;{\mathcal G}_S) > I(y;\mathcal{G}^k_S) + I(y;\mathcal{G}^{\backslash k}_S)$ follows for any ${\mathcal G}^{k}_S \in {\mathcal G}_S$.

\qed

For further analysis of Lemma 1, see Appendix~\ref{Assumption of Non-Independence}.~We discuss subgraph connectivity in Appendix~\ref{ax:Connectivity}.
\subsection{Synergistic Subgraphs Extraction}
Lemma~1 establishes the inequality properties of synergistic subgraphs. Following Lemma~1, we theoretically formulate an optimization problem for extracting the set of synergistic subgraphs.
Specifically, we formally define the mask optimization problem for extracting the synergistic subgraph set as follows:
\begin{equation}
\mathop{\arg\max}\limits_{\mathbf{M}} \; I(y;\mathcal{G}_S) - \sum\nolimits_{k=1}^K I(y;\mathcal{G}^k_S)
,
\label{eq:mask optimization problem}
\end{equation}
where $\mathbf{M} = \left[ \mathbf{M}^1,...,\mathbf{M}^K \right]$ denotes the concatenated matrix of edge mask matrices across $K$ views.
We optimize masks $\mathbf{M}$ to maximize the mutual information between $\mathcal{G}_S$ and $y$, while minimizing the sum of the mutual information between $\mathcal{G}^k_S$ and $y$ for each view.

Furthermore, we introduce the following lemma to reformulate the maximization of the mutual information term as the minimization of conditional entropy.
\paragraph{Lemma 2.}Assuming that $\mathcal{G}_S$ contains all multi-view information, the optimization problem in Formula~\ref{eq:mask optimization problem} can be equivalently expressed as:
\begin{equation}
\mathop{\arg\min}\limits_{\mathbf{M}} \; p(y|\mathcal{G}_S)\log p(y|\mathcal{G}_S) - \sum\nolimits_{k=1}^K p(y|\mathcal{G}^k_S) \log p(y|\mathcal{G}^k_S) 
.
\label{eq:conditional entropy}
\end{equation}
\paragraph{Proof.}
\[
\begin{aligned}
&\mathop{\arg\max}\limits_{\mathbf{M}} \; I(y;\mathcal{G}_S) - \sum\nolimits_{k=1}^K I(y;\mathcal{G}^k_S) \\
 \Leftrightarrow & \mathop{\arg\max}\limits_{\mathbf{M}}  H(y) - H(y|\mathcal{G}_S) - \sum\nolimits_{k=1}^K \Big[ H(y) - H(y|\mathcal{G}^k_S) \Big]\\
\Leftrightarrow & \mathop{\arg\max}\limits_{\mathbf{M}} \; (1-k)H(y) - H(y|\mathcal{G}_S) + \sum\nolimits_{k=1}^K H(y|\mathcal{G}^k_S) \\
\Leftrightarrow & \mathop{\arg\max}\limits_{\mathbf{M}} \; -H(y|\mathcal{G}_S) + \sum\nolimits_{k=1}^K H(y|\mathcal{G}^k_S)\\
\Leftrightarrow &\mathop{\arg\min}\limits_{\mathbf{M}} \; H(y|\mathcal{G}_S) - \sum\nolimits_{k=1}^K H(y|\mathcal{G}^k_S) \\
\Leftrightarrow &\mathop{\arg\min}\limits_{\mathbf{M}} \; p(y|\mathcal{G}_S)\log p(y|\mathcal{G}_S) - \sum\nolimits_{k=1}^K p(y|\mathcal{G}^k_S) \log p(y|\mathcal{G}^k_S) 
\end{aligned}
\]
\qed

Lemma 2 transforms the mutual information optimization problem into a conditional entropy optimization problem, enabling the direct use of the cross-entropy loss function to optimize masks.

\begin{figure*}
  \centering
  \includegraphics[width=0.9\linewidth]{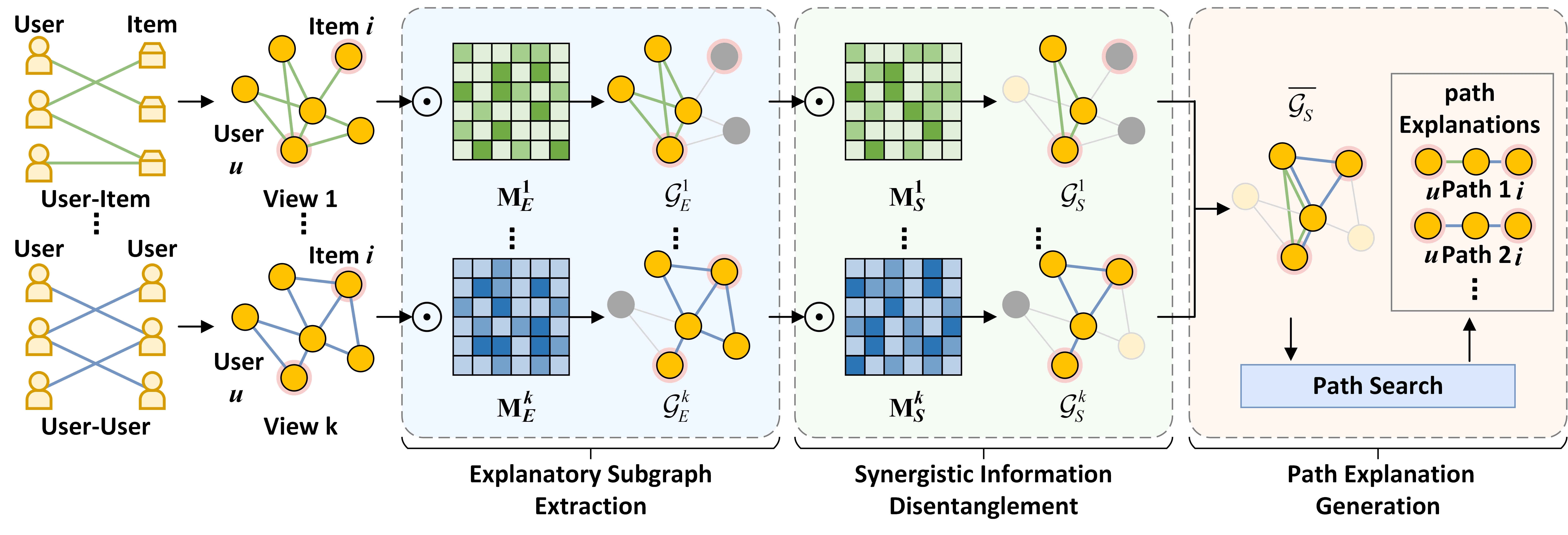}
      \vspace{-1em}
  \caption{Overall structure of SemExplainer. }
  \vspace{-1em}
  \Description{A woman and a girl in white dresses sit in an open car.}
  \label{fig:SemExplainer}
\end{figure*}

\section{Framework of SemExplainer}
In this section, we introduce the design details of SemExplainer. SemExplainer mainly consists of three core components: explanatory subgraph extraction, synergistic information disentanglement, and path explanation generation.
First, we utilize mask optimization to generate explanatory subgraphs in $K$ views for the recommendation system, which contain key topological information that influences recommendations.
Second, we exploit synergistic subgraph extraction theory to divide the explanatory subgraphs into two parts: synergistic subgraphs and non-synergistic subgraphs, thereby disentangling synergistic information from the explanatory subgraphs.
Finally, we search for paths from the user to the target item on the synergistic subgraph set as explanations for the synergistic effects between different views.
Fig.~\ref{fig:SemExplainer} shows the architecture of SemExplainer.

\subsection{Explanatory Subgraph Extraction}\label{section5.1}
In the social recommendation task, each type of social topological information, such as user-item interactions and user-user social relations, is input into the multi-view GNN as a single view graph.
The multi-view GNN integrates topological information from all views and predicts the similarity between the target user $u$ and the recommended item $i$.
This process is represented as follows:
\begin{equation}
{y} = f( {u,i,\mathcal{G}} )
,
\end{equation}
where $f(\cdot)$ denotes the social recommendation model based on multi-view GNN.
In this process, we consider topological information that generates synergistic effects as the key information influencing the recommendation.
Therefore, we employ mask learning to extract the key subset $\mathcal{G}_E^k$ of the topological information network $\mathcal{G}^k$ from $k$-th view that influences recommendations. 
Specifically, the adjacency matrix of $\mathcal{G}_E^k$ is represented as ${\mathbf{M}^k_E} \odot {\mathbf{A}^k}$, where $\mathbf{M}^k_E$ is the mask matrix, used to extract the explanatory subgraph of the $k$-th view.
To mitigate the distribution drift between $\mathcal{G}_E^k$ and $\mathcal{G}^k$, we adopt a soft mask strategy, defined as $\mathbf{M}^k_E \in [0,1]^{n \times n}$.
Let $\mathbf{M}^k_E$ follow the normal distribution, i.e., $m_E \sim \mathcal{N}(\mu  = 1.0,{\sigma ^2} = 2/n)$ for $m_e \in \mathbf{M}^k_E$.
To optimize all views of $\mathbf{M}_E=[\mathbf{M}^1_E,...,\mathbf{M}^K_E]$, we input graph set $\mathcal{G}_E=\{\mathcal{G}_E^1,...,\mathcal{G}_E^K\}$ into the social recommendation model:
\begin{equation}
{y}_E = f( {u,i,\mathcal{G}_E} )
,
\end{equation}
where ${y}_E$ represents the prediction of the recommendation generated by the input $\mathcal{G}_E$.
We minimize the following objective to maximize the mutual information between $\mathcal{G}_E$ and $y$:
\begin{equation}
\mathcal{L}_{E}(\mathbf{M}_E) = l(y,y_E)+\lambda  \cdot \sigma (\sum\nolimits_{k = 1}^K {{{\left\| \mathbf{M}^k_E \right\|}_1}} )
,
\end{equation}
where $l(\cdot)$ denotes the loss of cross-entropy. 
Note that $\mathcal{L}_{E}$ is optimized with respect to $\mathbf{M}_E$ only, while the parameters of $f(\cdot)$ are frozen.
In addition, we constrain the sparsity of $\mathcal{G}_E$ by minimizing a regularization term $\sigma (\sum\nolimits_{k = 1}^K {{{\left\| \mathbf{M}^k_E \right\|}_1}} )$, where $\sigma(\cdot)$ denotes the sigmoid function and $\left\| \cdot \right\|_1$ computes the $l_1$-norm.
$\lambda$ is a hyperparameter that controls the sparsity of the explanatory subgraphs.
The optimized $\mathcal{G}_E$ is used as a set of explanatory subgraphs to provide preliminary explanations for the social recommendation model.
\subsection{Synergistic Information Disentanglement}
To explain the synergistic effects across different views, it is necessary to further disentangle the synergistic information from the explanatory subgraph extracted in Section~\ref{section5.1}.
Specifically, we divide the explanatory subgraph $\mathcal{G}_E^k$ of the $k$-th view into two parts: the synergistic subgraph $\mathcal{G}_S^k$ and the non-synergistic subgraph $\mathcal{G}_{NS}^k$.
In theory, synergistic and non-synergistic information are mutually exclusive. For example, if an edge in the graph is not part of a synergistic subgraph, it must belong to a non-synergistic subgraph.
Therefore, given a mask optimized for capturing synergistic subgraphs, its inverse can naturally be used to extract non-synergistic subgraphs.
Similar to Section~\ref{section5.1}, we denote the adjacency matrix of 
$\mathcal{G}_S^k$ as $\mathbf{M}^k_S\odot\mathbf{M}^k_E\odot\mathbf{A}^k$ and that of $\mathcal{G}_{NS}^k$ as $(1-\mathbf{M}^k_{S})\odot\mathbf{M}^k_E\odot\mathbf{A}^k$, where $\mathbf{M}^k_S$ is the soft mask matrix employed to disentangle $\mathcal{G}_S^k$ and $\mathcal{G}_{NS}^k$.

Following the optimization problem formulated in Formula~\ref{eq:mask optimization problem}, we maximize the mutual information between the set $\mathcal{G}_S$ of all synergistic subgraphs and $y$, while minimizing the sum of mutual information between each individual synergistic subgraph $\mathcal{G}_S^k$ from different views and $y$. This process effectively maximizes the topological information gain derived from combining synergistic subgraphs across multiple views.
Specifically, we input $\mathcal{G}_S$ and $\mathcal{G}_S^k$ from the $k$-th view into the social recommendation model, respectively.
\begin{equation}
{y}_{S}^k = f({u,i,\mathcal{G}_S^k})
,\;\;\;\;\;\;
{y}_{S} = f({u,i,\mathcal{G}_S})
,
\label{eq:input SR model}
\end{equation}
where ${y}_{S}^k$ is the recommendation prediction based on $\mathcal{G}_S^k$, and ${y}_{S}$ is the prediction based on $\mathcal{G}_S$.
Furthermore, following the conditional entropy optimization problem derived from Formula~\ref{eq:conditional entropy}, we minimize the following objective to optimize $\mathbf{M}_S=[\mathbf{M}^1_S,...,\mathbf{M}^K_S]$:
\begin{equation}
\mathcal{L}_{S}(\mathbf{M}_S) = l(y,{y}_{S})-\sum\nolimits_{k = 1}^K {l(y,{y}_{S}^k)}
.
\end{equation}
It is worth noting that $\mathcal{L}_{S}$ optimizes only $\mathbf{M}_S$ while keeping $\mathbf{M}_E$ fixed, thereby ensuring that $\mathbf{M}_E$ consistently captures explanatory subgraphs.

By optimizing $\mathcal{L}_{S}$, we can extract information from explanatory subgraphs that sufficiently capture synergistic effects.
However, we consider the necessity of synergistic subgraphs.
In other words, synergistic subgraphs should exclusively contain synergistic information while excluding non-synergistic information, thereby achieving a disentanglement between the two.
To this end, we constrain the non-synergistic subgraph set $\mathcal{G}_{NS}$ to contain non-synergistic information. 
Similar to Formula~\ref{eq:input SR model}, we input $\mathcal{G}_{NS}^k$ and $\mathcal{G}_{NS}$ into the social recommender system and obtain two recommendation predictions ${y}_{NS}^k$ and ${y}_{NS}$.
\begin{equation}
{y}_{NS}^k = f({u,i,\mathcal{G}_{NS}^k})
,\;\;\;\;\;\;
{y}_{NS} = f({u,i,\mathcal{G}_{NS}})
.
\end{equation}
Subsequently, we optimize the inverse mask of $\mathbf{M}_S$ by minimizing the following objective in order to extract non-synergistic subgraphs:
\begin{equation}
\mathcal{L}_{NS}(\mathbf{M}_S) = \sum\nolimits_{k = 1}^K {l(y,{y}_{NS}^k)} - l(y,{y}_{NS})
.
\end{equation}
It is worth noting that the optimization objective of $\mathcal{L}_{NS}$ is precisely the opposite of that of $\mathcal{L}_{S}$.
\paragraph{Joint Optimization.}
Overall, we jointly optimize two mask matrices $\mathbf{M}_E$ and $\mathbf{M}_S$ to extract explanatory subgraphs and synergistic subgraphs for all views, respectively. The overall objective is as follows:
\begin{equation}
\mathcal{L}_{ALL}(\mathbf{M}_E,\mathbf{M}_S) = \mathcal{L}_{E}+\mathcal{L}_{S}+\mathcal{L}_{NS}
.
\end{equation}
We input a single set of multi-view topological information networks into the explainer and iteratively optimize $\mathcal{L}_{ALL}$ to obtain the explanatory subgraph set $\mathcal{G}_{E}$ and the synergistic subgraph set $\mathcal{G}_{S}$. 
Thus, this process constitutes optimization rather than learning, and during this procedure, the explainer does not access or modify any parameters of the social recommendation model.
In addition, we emphasize that although the optimization objectives for $\mathbf{M}_S$ and $\mathbf{M}_E$ differ, their optimization directions are consistent. 
The process of capturing explanatory subgraphs via $\mathbf{M}_E$ essentially performs a coarse-grained selection of synergistic information, as synergistic information exists exclusively in topological information that influences recommendation predictions.

\subsection{Path Explanation Generation}
The path connecting the target user and the recommended item provides a concise and interpretable explanation of the recommendation result. 
In social recommendation tasks, the recommendation prediction can be viewed as a link prediction problem, where the goal is to infer a preference-type edge between the user node and the item node.
Therefore, the path connecting the target user and the recommended item can naturally serve as an explanation for the link prediction between them. 
For instance, the recommendation of item $i_2$ to user $u_1$ can be justified by the fact that $u_1$ previously purchased item $i_1$, which shares the attribute $a_1$ with $i_2$. 
In this case, $\mathcal{P}=\{(u_1,i_1),(i_1,a_1),(a_1,i_2)\}$ represents the path, where $(u_1,i_1)$ denotes a directed edge from $u_1$ to $i_1$, and all other edges are defined analogously.

In this paper, we leverage paths to explain the synergistic effects in social recommendation.
We integrate the synergistic subgraphs from all views to construct a heterogeneous graph $\overline {\mathcal{G}_S} = (\mathcal{V}_S, {\cup _{k \in \{1,..., K\}}}\{ \mathcal{E}_S^k\})$.
We search for paths from target user $u$ to recommended item $i$ on $\overline {\mathcal{G}_S}$, constraining these paths to cover edges with high synergy effects. 
This ensures that the selected paths adequately capture and explain the synergistic effects underlying the recommendation prediction.
Specifically, for each edge $e$, we measure the element-wise product of its corresponding mask values, denoted as $m_e \in \mathbf{M}_E\odot\mathbf{M}_S$.
The larger the $m_e$ value, the richer the synergistic information encoded in the edge $e$, and the greater its impact on recommendation prediction.
Therefore, we define a synergistic effect score for each edge in $\overline {\mathcal{G}_S}$ as shown below:
\begin{equation}
S_e = \log {\sigma ( {m_e} )},\quad e \in {\cup _{k \in \{1,...,K\}}}\{ \mathcal{E}_S^k\},
\,m_e \in \mathbf{M}_E\odot\mathbf{M}_S
\label{eq}
,
 \end{equation}
where $S_e$ denotes the synergistic effect score of edge $e$.
Furthermore, we calculate the total score for a path as shown below:
\begin{equation}
S_\mathcal{P} = \sum\nolimits_{e \in \mathcal{P}} {S_e}
\label{eq:scorefunction}
,
 \end{equation}
where $S_\mathcal{P}$ denotes the synergistic effect score of path $\mathcal{P}$.
After calculating $S_e$ for each edge, we employ Dijkstra's~\cite{yucage} algorithm to search for paths that maximize $S_\mathcal{P}$. 
Since explanations are not necessarily unique, multiple paths may correspond to a single recommendation prediction. 
We select the top-$k$ paths with the highest $S_\mathcal{P}$ as the final explanations.

\section{Experiments and Results}

\begin{table*}[t]
  \centering
  \caption{Comparison on the ACM and Last-FM datasets~(\%).~$\uparrow$ indicates that higher values are better; $\downarrow$ indicates the opposite. The optimal result is highlighted in bold.}
    \begin{tabular}{c|l|cccc|cccc}
\toprule
\multirow{2}[4]{*}{Categories} & \multicolumn{1}{c|}{\multirow{2}[4]{*}{Methods}} & \multicolumn{4}{c|}{ACM}      & \multicolumn{4}{c}{Last-FM} \\
\cmidrule{3-10}          &       & FID+ $\uparrow$  & FID- $\downarrow$  & SIS $\uparrow$   & SIN $\downarrow$   & FID+ $\uparrow$  & FID- $\downarrow$  & SIS $\uparrow$   & SIN $\downarrow$ \\
    \midrule
    \multirow{5}[2]{*}{GNN-generic} & GNNExplainer & 94.11$\pm$1.2 & 62.18$\pm$2.4 & 32.77$\pm$1.7 & 70.59$\pm$4.8 & 81.10$\pm$0.1 & 51.00$\pm$0.1 & 65.33$\pm$1.4 & 64.52$\pm$1.2 \\
          & PGExplainer & 90.21$\pm$1.4 & 43.68$\pm$4.1 &17.54$\pm$3.2 & 58.62$\pm$9.2 & 66.90$\pm$2.1 & 49.01$\pm$10.3 & 48.81$\pm$2.8 & 71.70$\pm$2.8 \\
          & AxiomLayeredge & 91.02$\pm$6.1 & 40.23$\pm$2.5 & 47.25$\pm$2.1 & 52.11$\pm$3.6 & 80.11$\pm$3.1 & 40.17$\pm$2.2 & 54.76$\pm$4.1 & 60.24$\pm$3.7 \\
          & MAGE  & 93.98$\pm$3.2 & 30.36$\pm$8.1 & 48.12$\pm$1.3 & 44.87$\pm$6.1 & 82.71$\pm$2.0 & 52.70$\pm$6.4 & 52.90$\pm$3.6 & 49.57$\pm$4.9 \\
          & GraphSHAP-IQ & \underline{\textbf{94.20$\pm$3.0}} & 40.22$\pm$5.2 & 51.34$\pm$1.2 & 49.22$\pm$3.1 & 76.29$\pm$2.3 & 42.12$\pm$4.1 & 44.71$\pm$2.9 & 64.86$\pm$4.1 \\
    \midrule
    \multirow{4}[2]{*}{system-specific} & PaGE-Link   & 87.92$\pm$3.2 & 67.13$\pm$1.7 & 40.65$\pm$3.1 & 54.23$\pm$1.2 & 86.53$\pm$1.3 & 28.20$\pm$1.3 & 35.85$\pm$1.0 & 79.65$\pm$2.9 \\
          & xPath   & 91.11$\pm$4.6 & 50.19$\pm$0.2 & 42.45$\pm$3.7 & 54.25$\pm$3.7 & 80.71$\pm$4.3 & 28.01$\pm$2.4 & 43.77$\pm$1.8 & 44.59$\pm$1.6 \\
          & SR-GCA   & 90.33$\pm$3.0 & 52.23$\pm$2.1 & 47.66$\pm$2.1 & 65.11$\pm$2.3 & 90.02$\pm$2.4 & 29.35$\pm$2.8 & 47.12$\pm$3.2 & 49.37$\pm$4.0 \\
          & CaGE  & 91.49$\pm$1.2 & 68.64$\pm$1.5 & 27.79$\pm$3.4 & 78.52$\pm$8.1 & 90.38$\pm$0.1 & 27.85$\pm$3.1 & 35.95$\pm$1.1 & 79.78$\pm$2.0 \\
    \midrule
    Ours  & SemExplainer & 91.81$\pm$2.7 & \underline{\textbf{24.55$\pm$5.8}} & \underline{\textbf{87.78$\pm$1.5}} & \underline{\textbf{14.03$\pm$2.0}} & \underline{\textbf{90.41$\pm$2.1}} & \underline{\textbf{18.13$\pm$4.1}} & \underline{\textbf{69.41$\pm$2.1}} & \underline{\textbf{40.87$\pm$3.7}} \\
    \bottomrule
    \end{tabular}%
  \label{tab:ex-Comparison Results 1}%
\end{table*}%

\subsection{Experimental Setup}
\subsubsection{Datasets and Baselines}
To evaluate the effectiveness of SemExplainer, we conduct experiments on three social recommendation datasets: AugCitation~\cite{10.1145/3637528.3672354}, Last-FM~\cite{10.1145/3605357}, and ACM~\cite{li2025clustering}. 
Additionally, we choose two categories of baseline methods, including
(1)~General GNN explainable methods:~GNNExplainer~\cite{ying2019gnnexplainer}, PGExplainer~\cite{luo2020parameterized}, AxiomLayeredge~\cite{liu2025explanations}, MAGE~\cite{yumage}, and GraphSHAP-IQ~\cite{muschalikexact}. 
These methods can explain general GNNs. 
(2)~RS explainable methods:~PaGE-Link~\cite{zhang2023page}, xPath~\cite{li2023towards}, SR-GCA~\cite{huang2025social}, and CaGE~\cite{yucage}. 
These methods focus on explaining GNNs in the social recommendation task.
More details on the datasets and baselines are provided in Appendix~\ref{ap:Datasets} and~\ref{ap:Baselines}.
\subsubsection{Evaluation Metrics}
We assess the validity of SemExplainer using the following evaluation metrics:
\begin{itemize}
\item {\textbf{Fidelity+ (FID+) and Fidelity- (FID-)}}~\cite{pmlr-v235-chen24d}:~FID+ quantifies the proportion of explanatory subgraphs that can successfully recover the original prediction. It is computed as follows:
\begin{equation}
\text{FID+} = {\mathbb{E}}[ {\mathbb{I}\left( y = f( {u,i,\mathcal{G}_E} )\right)} ]
,
\end{equation}
where $\mathbb{I}(\cdot)$ denotes the indicator function, defined as $\mathbb{I}(\cdot)=1$ if and only if the condition holds, otherwise $\mathbb{I}(\cdot)=0$.
FID- quantifies the proportion of non-explanatory subgraphs that can recover the original prediction. It is computed as:
\begin{equation}
\text{FID-} = {\mathbb{E}}[ {\mathbb{I}\left( y = f( {u,i,\mathcal{G}\backslash \mathcal{G}_E} )\right)} ]
,
\end{equation}
where $\mathcal{G}\backslash \mathcal{G}_E$ denotes the non-explanatory subgraph, defined as the difference between the original graph and the explanatory subgraph.
\item {\textbf{Synergistic Interaction Score~(SIS) and Non-Synergistic Interaction Score~(SIN)}}~\cite{kim2024discovering}:~SIS is the proportion of test samples for which the multi-view synergistic subgraph provides greater mutual information with $y$ than the sum of its individual view components.
SIS quantifies the information gain ratio contributed by synergistic topological information. 
Higher SIS values indicate stronger synergistic effects. SIS is computed as follows:
\begin{equation}
\text{SIS} = {\mathbb{E}}[ {\mathbb{I}( {I(y,{\mathcal{G}_S}) > \sum\nolimits_{k = 1}^K {I(y,\mathcal{G}_S^k)} } )} ]
.
\end{equation}
SIN is the opposite of SIN. SIN quantifies the information attenuation ratio of non-synergistic subgraph:
\begin{equation}
\text{SIN} = {\mathbb{E}}[ {\mathbb{I}( {I(y,{\mathcal{G}_{NS}}) \ge  \sum\nolimits_{k = 1}^K {I(y,\mathcal{G}_{NS}^k)} } )} ]
.
\end{equation}
A lower SIN value indicates that less synergistic topological information is present in $\mathcal{G}_{NS}$.
\item {\textbf{Sparsity (SPA)}}~\cite{pmlr-v235-chen24d}:~SPA quantifies the proportion of edges in the explanatory subgraph relative to the total number of edges in the original input graph. It is computed as follows:
\begin{equation}
\text{SPA} = {\mathbb{E}}[1 - \frac{{\left| {\mathcal E}_E \right|}}{{\left| \mathcal E \right|}}]
,
\end{equation}
where ${\mathcal E}$ represents the number of edges in the original graph, and ${\mathcal E}_E$ represents the number of edges in $\mathcal{G}_E$.
\end{itemize}

\subsection{Comparison Results}
\begin{table}[t]
  \centering
  \caption{Comparison on the AugCitation dataset~(\%).~$\uparrow$ indicates that higher values are better; $\downarrow$ indicates the opposite. The optimal result is highlighted in bold.}
        \begin{tabularx}{1\linewidth}
    {>{\arraybackslash}m{2.3cm}|
    >{\centering\arraybackslash}m{1.1cm}
    >{\centering\arraybackslash}m{1.1cm}
    >{\centering\arraybackslash}m{1.1cm}
    >{\centering\arraybackslash}m{1.1cm}
    >{\centering\arraybackslash}m{1.1cm}
  }
\toprule    
Methods & FID+ $\uparrow$  & FID- $\downarrow$  & SIS $\uparrow$   & SIN $\downarrow$ \\
    \midrule
    GNNExplainer & 67.70$\pm$3.6 & 14.76$\pm$2.3 & 53.11$\pm$3.2 & 49.71$\pm$2.1 \\
    PGExplainer & 40.04$\pm$0.1 & 31.76$\pm$0.1 & 52.33$\pm$4.0 & 62.20$\pm$3.2 \\
    AxiomLayeredge & 70.21$\pm$4.1 & 14.80$\pm$2.1 & 45.28$\pm$5.3 & 72.09$\pm$6.3 \\
    MAGE  & 70.28$\pm$2.2 & 21.11$\pm$1.1 & 49.71$\pm$2.9 & 48.83$\pm$3.3 \\
    GraphSHAP-IQ & 69.92$\pm$3.6 & 14.6$\pm$4.1 & 32.10$\pm$7.2 & 57.13$\pm$4.4 \\
    \midrule
    PaGE-Link   & 57.82$\pm$2.4 & 11.10$\pm$3.7 & 52.15$\pm$7.1 & 65.03$\pm$2.6 \\
    xPath   & 51.63$\pm$2.1 & 35.32$\pm$1.7 & 60.33$\pm$5.2 & 54.12$\pm$5.0 \\
    SR-GCA   & 71.12$\pm$2.0 & 12.03+3.6 & 61.24$\pm$5.3 & 53.72$\pm$3.9 \\
    CaGE  & 70.81$\pm$1.3 & 11.74$\pm$3.9 & 34.10$\pm$5.2 & 68.09$\pm$2.0 \\
    \midrule
    SemExplainer & \underline{\textbf{71.56$\pm$1.1}} & \underline{\textbf{10.99$\pm$4.1}} & \underline{\textbf{96.40$\pm$1.3}} & \underline{\textbf{10.07$\pm$1.7}} \\
    \bottomrule
    \end{tabularx}%
  \label{tab:ex-Comparison Results 2}%
\end{table}%

To evaluate the effectiveness of SemExplainer in explaining synergistic effects, we compare it with eight baseline methods on three datasets. 
We report FID+, FID-, SIS, and SIN for all methods in Tables~\ref{tab:ex-Comparison Results 1} and~\ref{tab:ex-Comparison Results 2}.
\begin{table}[h]
  \centering
  \caption{Comparison of sparsity on the three datasets~(\%).}
    \begin{tabularx}{1\linewidth}
    {>{\arraybackslash}m{2.3cm}|
    >{\centering\arraybackslash}m{1.56cm}
    >{\centering\arraybackslash}m{1.56cm}
    >{\centering\arraybackslash}m{1.56cm}
  }
    \toprule
    \multicolumn{1}{c|}{Methods} & ACM   & Last-FM & AugCitation \\
    \midrule
    GNNExplainer & 97.56$\pm$1.8 & 59.23$\pm$7.7 & 80.49+9.1 \\
    PGExplainer & 93.36$\pm$1.4 & 73.40$\pm$1.6 & 63.36$\pm$8.3 \\
    AxiomLayeredge & 99.12$\pm$0.2 & 98.65$\pm$0.7 & 98.25$\pm$1.0 \\
    MAGE  & 98.70$\pm$1.1 & 81.92$\pm$3.5 & 90.77$\pm$1.2 \\
    GraphSHAP-IQ & 91.02$\pm$6.4 & 88.23$\pm$3.1 & 89.70$\pm$5.1 \\
    \midrule
    PaGE-Link   & 99.01$\pm$0.3 & 83.91$\pm$4.2 & 82.74$\pm$2.8 \\
    xPath   & 99.71$\pm$0.1 & 82.21$\pm$6.3 & 81.46$\pm$1.9 \\
    SR-GCA   & 99.12$\pm$0.6 & 97.46$\pm$1.1 & 96.10$\pm$2.2 \\
    CaGE  & 98.59$\pm$0.1 & 84.65$\pm$4.2 & 89.15$\pm$1.5 \\
    \midrule
    SemExplainer-GE & 98.72$\pm$0.9 & 98.70$\pm$0.1 & 98.44$\pm$0.3 \\
    SemExplainer-GS & 99.02$\pm$0.7 & 98.81$\pm$0.2 & 98.61$\pm$0.1 \\
    SemExplainer-PA & \underline{\textbf{99.78$\pm$0.1}} & \underline{\textbf{99.02$\pm$0.0}} & \underline{\textbf{98.67$\pm$0.1}} \\
    \bottomrule
    \end{tabularx}%
  \label{tab:Comparison of sparsity on the three datasets}%
  \vspace{-1em}
\end{table}%
The results demonstrate that SemExplainer consistently achieves the best SIS and SIN scores across all three datasets, indicating that it explains synergistic effects in multi-view GNNs more effectively than state-of-the-art baselines. 
Meanwhile, SemExplainer attains comparable FID+ and FID- scores to the baselines, suggesting that while capturing synergistic effects, it also faithfully accounts for the factors influencing recommendation. 
This further demonstrates that the synergistic topological information extracted by SemExplainer constitutes key determinants of recommendation predictions.

Additionally, we use the SPA metric to evaluate the sparsity of explanations across the three datasets. 
A higher SPA value indicates more concise and interpretable explanations. 
For a complete assessment, Table~\ref{tab:Comparison of sparsity on the three datasets} reports the sparsity of SemExplainer-generated explanatory subgraphs (labeled 'SemExplainer-GE'), synergistic subgraphs (labeled 'SemExplainer-GS'), and paths (labeled 'SemExplainer-PA').
The results show that SemExplainer achieves sparsity comparable to the baselines, indicating that it maintains explanation readability while capturing synergistic effects. 
Notably, the sparsity of SemExplainer-generated path explanations and synergistic subgraphs exceeds that of general explanatory subgraphs across all three datasets. 
This is expected, as synergistic information is inherently a subset of explanatory information and is therefore naturally sparser.

\begin{figure*}[h]
  \centering
  \includegraphics[width=0.9\linewidth]{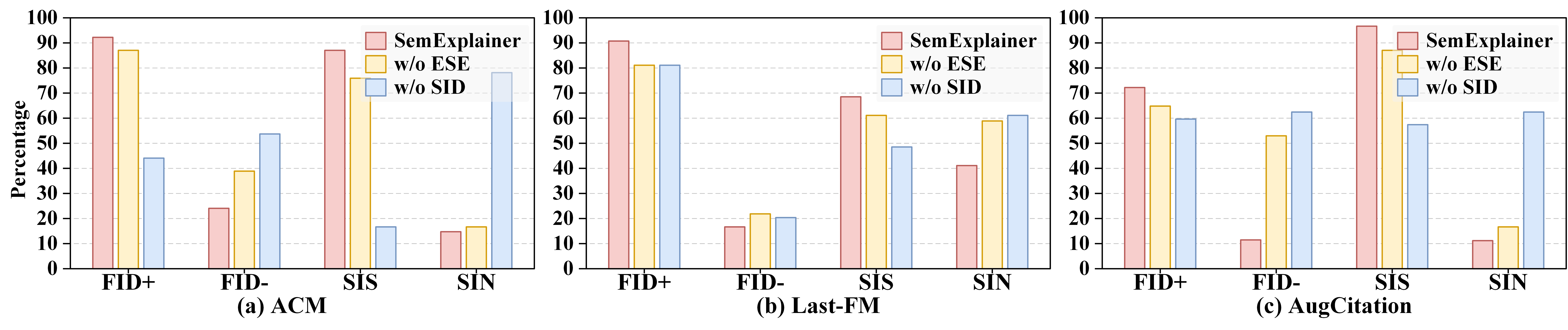}
    \vspace{-1em}
\caption{Ablation analysis across three datasets. ‘w/o ESE’~denotes removal of the explanatory subgraph extraction component, ‘w/o SID’~denotes removal of the synergistic information disentanglement component.}
\vspace{-1em}
  \Description{A woman and a girl in white dresses sit in an open car.}
  \label{ex: Ablation Study}
\end{figure*}
\vspace{-1em}
\subsection{Ablation Studies}

\begin{figure*}[h]
  \centering
  \includegraphics[width=0.9\linewidth]{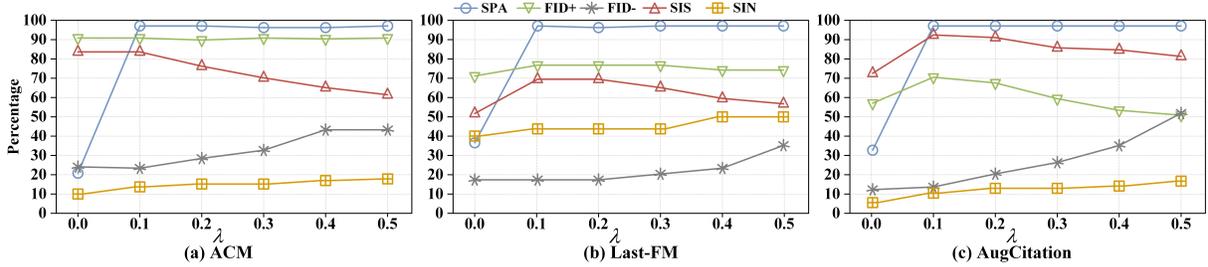}
\vspace{-1em}
\caption{The effect of hyperparameter $\lambda$ on metrics FID+, FID-, SIS, SIN, and SPA.}
\vspace{-1em}
  \Description{A woman and a girl in white dresses sit in an open car.}
  \label{ex: hyperparameter}
\end{figure*}

To assess the necessity of components within SemExplainer, we conduct ablation analyses on several key modules. 
Specifically, the explanatory subgraph extraction component~(ESE) and the synergistic information disentanglement component~(SID) are removed individually to evaluate their contributions. 
Figure~\ref{ex: Ablation Study} illustrates the impact of these removals on explaining synergistic effects. 
The results indicate that SemExplainer achieves state-of-the-art performance when all components are incorporated. 
Removal of the ESE component prevents the explainer from extracting meaningful explanatory information, resulting in decreased fidelity and non-synergistic subgraphs containing noise unrelated to recommendation predictions. 
Similarly, removing the SID component prevents the explainer from extracting synergistic information from explanatory subgraphs, resulting in poorer SIS and SIN scores. 
The removal of the path explanation generation component is not evaluated, as it does not affect the ability to explain synergistic effects; its role is limited to searching for paths to ensure explanations are concise and readable.

\subsection{Hyperparameter Analysis}

The SemExplainer model incorporates a hyperparameter $\lambda$ that governs the sparsity of the explanatory subgraph. 
Larger values of $\lambda$ impose stronger sparsity constraints in loss $\mathcal{L}_{E}$, whereas smaller values relax these constraints. 
We investigat the effect of $\lambda$ on explanatory fidelity, sparsity, and the capability to capture synergistic effects across three datasets, varying $\lambda$ from 0 to 1. 
As shown in Figure~\ref{ex: hyperparameter}, increasing $\lambda$ leads to decreases in FID+ and SIS, while FID- and SIN increase. 
This suggests that imposing stronger sparsity constraints slightly reduces the effectiveness of SemExplainer in explaining synergistic effects, as the mask must filter out weaker synergistic information to enhance sparsity. 
Correspondingly, higher $\lambda$ values also increase SPA. 
Based on these observations, we set $\lambda$ = 0.1 to balance sparsity and the ability to capture synergistic effects.

\section{CONCLUSION AND DISCUSSION}
In this paper, we explore how to explain the synergistic effects of multi-view GNNs in social recommendation tasks and propose SemExplainer, a multi-view GNN explainer.
Inspired by information gain quantification, we introduce the synergistic subgraph theory to support the identification of synergistic topological information.
Building on this foundation, SemExplainer leverages synergistic subgraph theory to formulate a conditional entropy optimization problem, thereby extracting synergistic topological information and generating path explanations for recommendations. 
It overcomes the inherent limitations of existing importance-based explainers by quantifying the information gain derived from synergistic effects. 
SemExplainer not only overcomes a key technical bottleneck in explainable multi-view GNNs but also offers a comprehensive explainability solution for social recommendation tasks, marking a substantial step toward bridging the trust gap between humans and web social platforms.
Additionally, our future work will focus on two limitations of SemExplainer: (1) it incurs higher computational overhead than baseline explainers, and (2) it is limited to generating instance-level explanations, lacking global explainability of synergistic effects. 
We leave two challenges for future work.

\section*{Acknowledgments}
This paper is supported by the National Natural Science Foundation of China (Grant No. 72204037) and the National Natural Science Foundation of China (Grant No. 62576075).

\bibliographystyle{ACM-Reference-Format}
\bibliography{sample-base}

\appendix
\section{Theoretical Analysis}
\subsection{Assumption of Non-Independence}\label{Assumption of Non-Independence}
Lemma 1 is predicated on the assumption that $\mathcal{G}^k_S$ and $\mathcal{G}^{\backslash k}_S$ are non-independent, as their synergistic relationship is fundamental to the formation of the synergistic subgraph.
This assumption is necessary; if $\mathcal{G}^k_S$ and $\mathcal{G}^{\backslash k}_S$ were independent, no synergistic effects would exist between them, and the following inequality would hold instead:
\begin{equation}
I(y;\mathcal{G}_S) \le I(y;\mathcal{G}^k_S) + I(y;\mathcal{G}^{\backslash k}_S)
\end{equation}
The above inequality contradicts the conclusion of Lemma 1. 
This contradiction demonstrates that the conditions of Lemma 1 for $\mathcal{G}^k_S$ and $\mathcal{G}^{\backslash k}_S$ are both sufficient and necessary; that is, the lemma holds if and only if synergistic effects exist between $\mathcal{G}^k_S$ and $\mathcal{G}^{\backslash k}_S$.
\subsection{Connectivity of the Synergistic Subgraph}\label{ax:Connectivity}
In graph theory definitions, subgraphs may be disconnected~\cite{bar2024flexible}. Therefore, during synergistic subgraph extraction, we similarly consider that synergistic subgraphs need not be connected. 
This insight aligns with intuition. 
In social recommendation systems, topological information reflecting synergistic effects typically connects either user nodes or item nodes, but not necessarily both simultaneously. 
Consequently, a user node may not be able to traverse a path from itself to an item node along the edge set of the synergistic topological information. 
This explains why SemExplainer maximizes the synergistic effect score when generating path explanations, without imposing a constraint that all edges in the explanation must belong to the synergistic subgraph.

\section{More Expertmental Results}

\subsection{User Studies}\label{ax:UserStudies}
\begin{figure}[h]
  \centering
  \includegraphics[width=\linewidth]{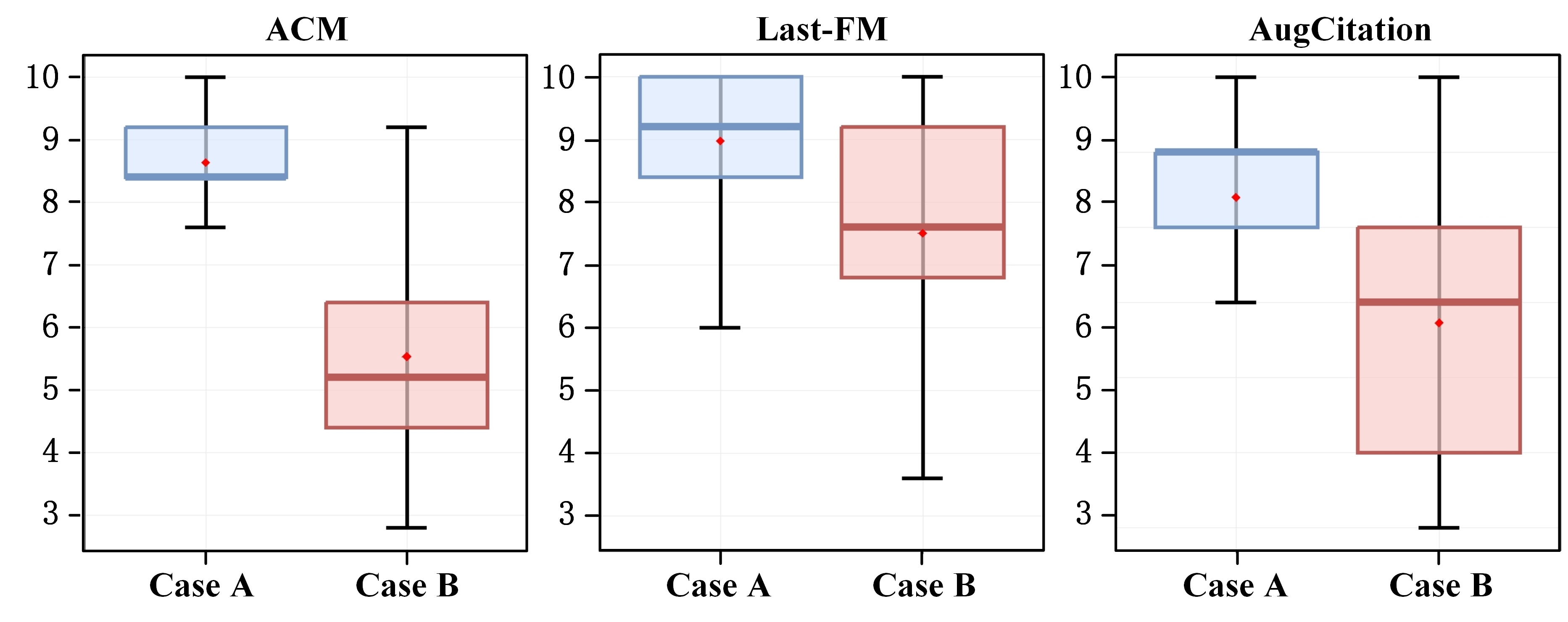}
\caption{Questionnaire results for the three datasets. The blue box plot depicts the score distribution for Case A, whereas the red box plot depicts the distribution for Case B. The red dot denotes the mean, and the two black lines indicate the maximum and minimum values.}
  \Description{A woman and a girl in white dresses sit in an open car.}
  \label{ex: userstudy}
\end{figure}

To assess the acceptance of recommendation explanations among end users, we conducted a user study using an A/B testing–based questionnaire survey. 
Explanations generated by SemExplainer and importance-based methods were designated as Case A and Case B, respectively, and users were asked to rate both. 
The rating criterion was whether the provided explanations enabled them to understand the synergistic effects underlying the recommendation results. Scores ranged from 0 (“completely unable to understand”) to 10 (“very easy to understand”).
The questionnaire was distributed to 600 end users. 
To ensure fairness, we balance the age distribution following the method proposed by Yu et al.~\cite{yucage}, defining users aged 20 and above as the adult cohort. 
Participants are evenly divided into six age groups (21–25, 26–30, 31–35, 36–40, 41–45, and 46–50), with 100 participants per group, and gender balance was maintained within each group. 
The survey was conducted across three datasets, yielding 511 valid responses. The response rate for the survey is 85.16\%.
As illustrated in Figure~\ref{ex: userstudy}, SemExplainer-generated explanations achieve higher average scores and lower variance compared to importance-based explanations. 
These results demonstrate that SemExplainer more effectively conveys the origins of synergistic effects, thereby enhancing users’ understanding of the underlying formation mechanisms.
\subsection{Efficiency Analysis}
To evaluate computational efficiency, we compare the runtime of SemExplainer with two lightweight baseline methods. 
Table~\ref{ex:Efficiency Analysis-1} reports both the runtime and the average growth rate for each method after 20, 60, and 100 epochs. 
The results show that SemExplainer incurs a slightly higher runtime than the baselines, primarily due to its additional SID module, which requires computing cross-entropy separately for all views. 
Nevertheless, as the number of epochs increases, the runtime growth rate of SemExplainer becomes comparable to that of the baselines. 
This demonstrates that although SemExplainer is more computationally demanding, it maintains good scalability.

\begin{table}[htbp]
  \centering
  \caption{Comparison of computational efficiency. $\Delta_{gr}$ indicates the average growth rate of runtime~(s) for 20, 60, and 100 epochs.}
    \begin{tabularx}{1\linewidth}
    {>{\arraybackslash}m{1.9cm}|
    >{\centering\arraybackslash}m{0.90cm}|
    >{\centering\arraybackslash}m{1.3cm}
    >{\centering\arraybackslash}m{1.3cm}
    >{\centering\arraybackslash}m{1.3cm}
  }
    \toprule
    \multicolumn{1}{c|}{Methods} & \multicolumn{1}{c|}{Epochs} & ACM   & Last-FM & AugCitation \\
    \midrule
    \multirow{4}[2]{*}{GNNExplainer} & 20    & 3.83$\pm$4.1 & 4.24$\pm$5.1 & 4.43$\pm$6.2 \\
          & 60    & 7.78$\pm$3.8 & 8.34$\pm$5.0 & 8.61$\pm$4.7 \\
          & 100    & 10.55$\pm$4.2 & 11.46$\pm$6.2 & 11.92$\pm$6.1 \\
          \cmidrule{2-5}
          & $\Delta_{gr}$    & 0.693 & 0.670 & 0.664 \\
    \midrule
    \multirow{4}[2]{*}{CaGE} & 20    & 5.73$\pm$1.2 & 7.57$\pm$2.0 & 8.11$\pm$4.7 \\
          & 60    & 12.10$\pm$1.7 & 14.02$\pm$2.3 & 15.36$\pm$6.1 \\
          & 100    & 13.19$\pm$4.2 & 15.73$\pm$5.1 & 15.92$\pm$5.1 \\
          \cmidrule{2-5}
          & $\Delta_{gr}$    & 0.601 & 0.487 & 0.465 \\
    \midrule
    \multirow{4}[2]{*}{SemExplainer} & 20    & 4.58$\pm$3.3 & 7.04$\pm$3.1 & 7.96$\pm$4.2 \\
          & 60    & 9.38$\pm$2.6 & 12.03$\pm$1.4 & 12.74$\pm$3.8 \\
          & 100    & 11.18$\pm$7.3 & 14.98$\pm$5.4 & 15.61$\pm$3.6 \\
          \cmidrule{2-5}
          & $\Delta_{gr}$    & 0.619 & 0.477 & 0.412 \\
    \bottomrule
    \end{tabularx}%
  \label{ex:Efficiency Analysis-1}%
\end{table}%

\subsection{Qualitative Analysis}
\begin{figure}
  \centering
  \includegraphics[width=\linewidth]{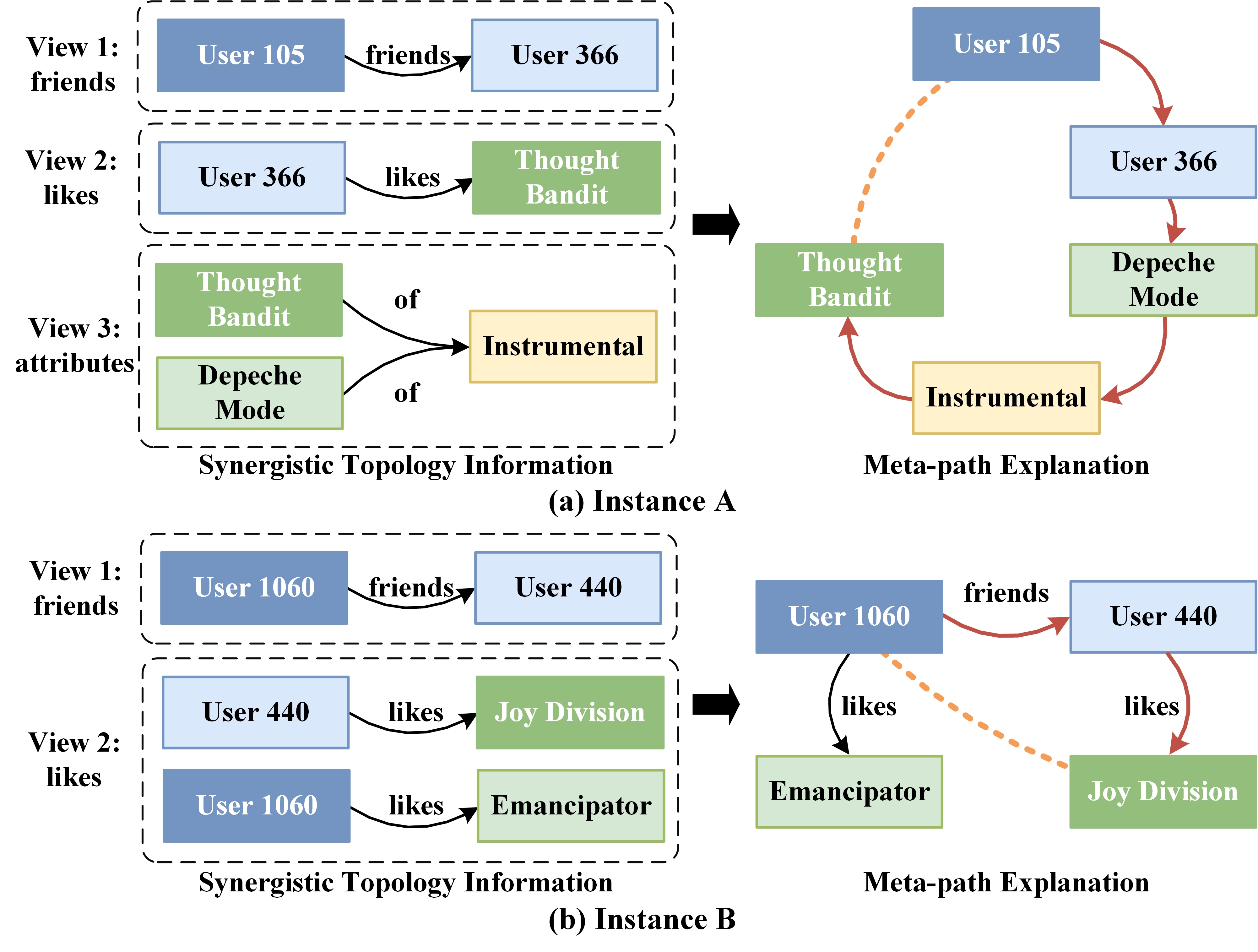}
\vspace{-1em}
\caption{Visualization of synergistic topology information and path explanation for two instances on the Last-FM dataset. The orange dashed line shows recommended actions, the red solid line represents the path, and the black solid line denotes the topological relationship.}
\vspace{-1.5em}
  \Description{A woman and a girl in white dresses sit in an open car.}
  \label{ex: Qualitative analysis}
\end{figure}

To assess the effectiveness and comprehensibility of SemExplainer in revealing synergistic effects, we visualize its generated explanations using the Last-FM dataset as an example. 
Figures~\ref{ex: Qualitative analysis}(a) and~\ref{ex: Qualitative analysis}(b) depict the synergistic topological information and the path explanations extracted by SemExplainer for two representative instances, respectively.
For instance A in Figure~\ref{ex: Qualitative analysis}(a), the synergistic information extracted by SemExplainer shows that recommending Artist Thought Bandit to User 105 arises from synergistic interactions across three views: user-user friendship connections, user-item preference interactions, and item-tag attribute relationships.
The path explanation for instance A illustrates the rationale behind the recommendation: User 105 is a friend of User 366, who likes the instrumental artist Depeche Mode. Since Thought Bandit is also an instrumental artist, the system recommends Thought Bandit to User 105.
Similarly, for instance B, the social recommender system leverages the synergistic effect between social relationships and preference interactions to generate the recommendation.
SemExplainer explains that the artist Joy Division is recommended to User 1060 because User 1060 is a friend of User 440, who likes the artist Joy Division.
The visualizations of the two instances demonstrate that the path explanations generated by SemExplainer enhance human understanding of the synergistic effects underlying recommendation decisions. 

\begin{figure*}
  \centering
  \includegraphics[width=\linewidth]{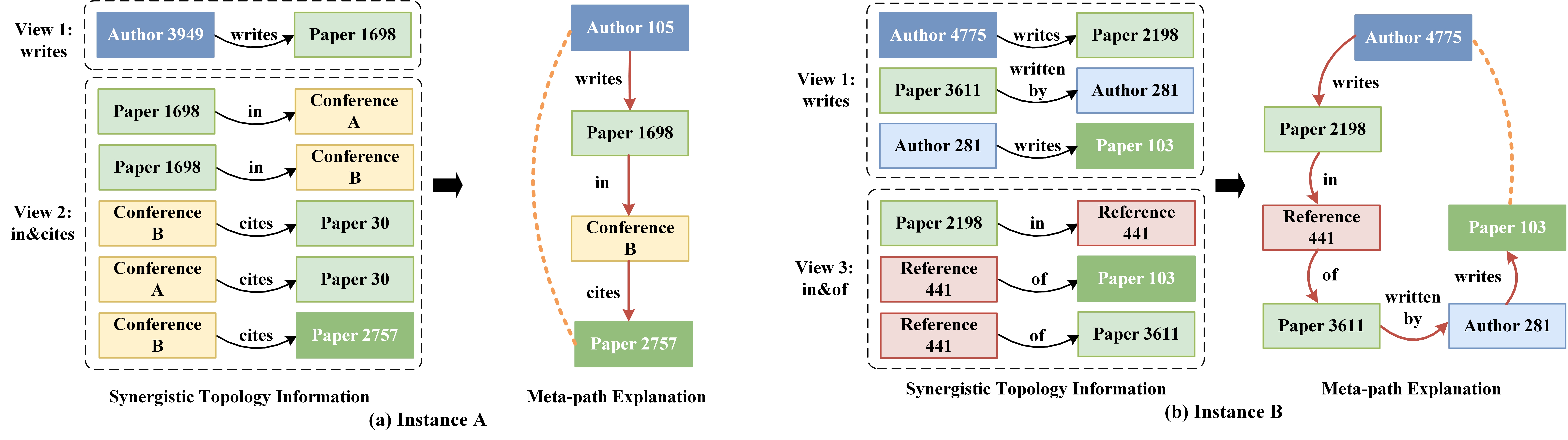}
  \caption{Visualization of synergistic topological information and path explanation for two instances on the AugCitation dataset. The orange dashed line shows recommended actions, the red solid line represents the path, and the black solid line denotes the topological relationship.}
  \Description{A woman and a girl in white dresses sit in an open car.}
  \label{fig:path2}
\end{figure*}

\begin{figure*}
  \centering
  \includegraphics[width=\linewidth]{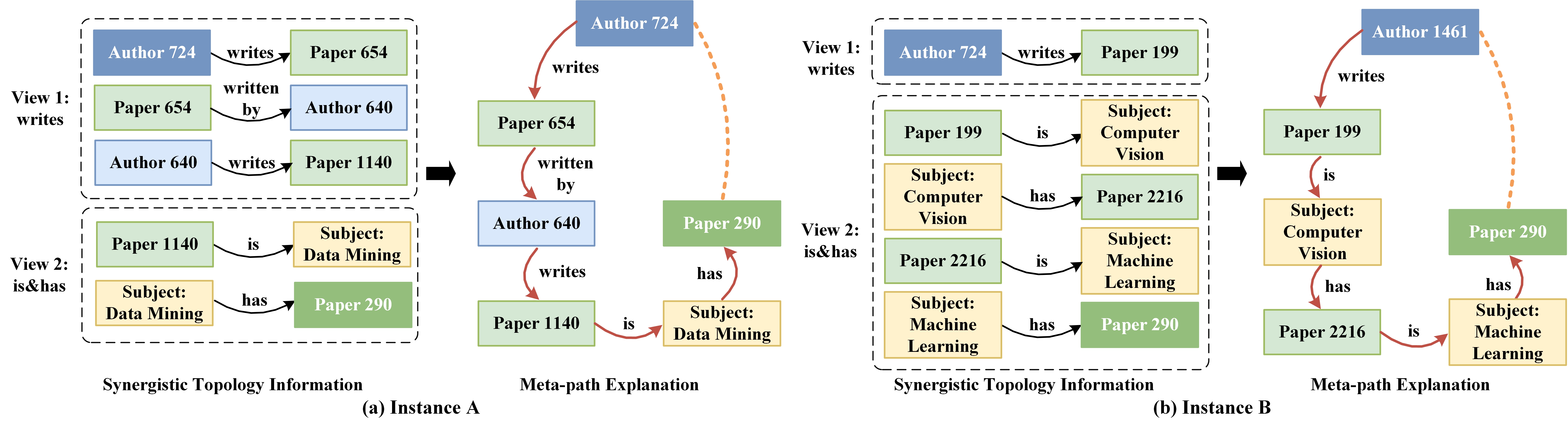}
  \caption{Visualization of synergistic topological information and path explanation for two instances on the ACM dataset. The orange dashed line shows recommended actions, the red solid line represents the path, and the black solid line denotes the topological relationship.}
  \Description{A woman and a girl in white dresses sit in an open car.}
  \label{fig:path3}
\end{figure*}

Figure~\ref{fig:path2} and Figure~\ref{fig:path3} visualize the synergistic topological information and path explanations for two instances on the ACM dataset and AugCitation dataset, respectively.
It should be noted that, owing to strict privacy protection policies governing authors and institutions in academic network data, only the subject attributes of papers are accessible. 
All other attributes are anonymized and represented numerically. 
Nevertheless, the visualized paths still allow for a meaningful interpretation of the synergistic interactions across different views.

\section{Expertmental Detatils}
\subsection{Datasets}\label{ap:Datasets}
Further details of the three datasets are provided as follows:
\begin{itemize}
\item {\textbf{ACM}}~dataset is an academic network consisting of three entity types: papers, authors, and subjects. It includes two types of relationships, namely author–paper and paper–subjects. We regard these relationships as distinct views and conduct the recommendation task of suggesting papers to authors.
\item {\textbf{Last-FM}}~dataset is a multi-view social network derived from the Last.fm online music platform. It consists of two entity types, users and artists, and three types of relationships: likes (user–artist), friends (user–user), and belongs-to (artist–user). We conduct the recommendation task of suggesting music artists to users.
\item {\textbf{AugCitation}}~dataset is constructed by augmenting the AMiner citation network~\cite{zhang2023page}. AugCitation is a multi-view citation network containing five entity types: authors, papers, paper tags, conferences, and references, as well as four relationship types: paper–tag, paper–author, paper–conference, and paper–reference. Each relationship is treated as a separate view. For the social recommendation task, we additionally introduce a “preference” (author–paper) relationship, which connects author entities to paper entities.
\end{itemize}
The detailed statistics of these datasets are summarized in Table~\ref{tab:Statistics of the Three Datasets}.
\begin{table}[h]
  \centering
  \caption{Statistics of the Three Datasets}
    \begin{tabularx}{1\linewidth}
    {>{\centering\arraybackslash}m{1.2cm}
    >{\centering\arraybackslash}m{1.2cm}|
    >{\centering\arraybackslash}m{1.5cm}
    >{\centering\arraybackslash}m{1.5cm}
    >{\centering\arraybackslash}m{1.5cm}
  }
    \toprule
    \multicolumn{2}{c|}{Dataset} & ACM&Last-FM&AugCitation \\
    \midrule
    \multirow{2}[1]{*}{Types} & $\#$Entity &    3   &    2   & 5 \\
          & $\#$Relation &   2    &   3    & 4 \\
    \midrule
    \multirow{2}[1]{*}{Numbers} & $\#$Entity &   11,246    &  31,469     & 439,457 \\
          & $\#$Relation &    23,664   &   475,410    & 1,002,845 \\
    \bottomrule
    \end{tabularx}%
  \label{tab:Statistics of the Three Datasets}%
\end{table}%
\subsection{Baselines}\label{ap:Baselines}
We provide detailed descriptions of two categories of baseline methods: GNN-general explainers and system-specific explainers.
\begin{itemize}
\item {\textbf{GNNExplainer}}~explains GNN predictions by learning masks and leveraging gradient computation to extract subgraphs that capture the most influential topological structures of the input graph.
\item {\textbf{PGExplainer}}~is capable of producing global explanations for GNN models by learning a unified explanatory subgraph that generalizes across multiple input graphs.
\item {\textbf{AxiomLayeredge}}~explains model predictions by analyzing how variations in edge weights within the input graph influence the output.
\item {\textbf{MAGE}}~decomposes the explanatory subgraph into multiple motifs, thereby enhancing both the interpretability and the practical significance of the explanation.
\item {\textbf{GraphSHAP-IQ}}~utilizes the Shapley value to explain the influence of interactions between nodes on GNN predictions.
\item {\textbf{PaGE-Link}}~generates paths to explain the link prediction between user nodes and item nodes in recommendation tasks.
\item {\textbf{xPath}}~searches for explanatory paths in heterogeneous graphs to interpret link predictions. In particular, it can be applied to elucidate the recommendation behaviors of social recommender systems.
\item {\textbf{SR-GCA}}~generates counterfactual explanations for the recommendations of the social recommender system. Moreover, it can leverage these counterfactual explanations to further improve recommendation performance.
\item {\textbf{CaGE}}~uses backdoor adjustment techniques to disentangle causal from non-causal relationships, generating causal path explanations for social recommender systems. It represents a state-of-the-art post-hoc explainability method specifically designed for social recommender systems.
\end{itemize}

\subsection{Implementation Details}\label{ap:ImplementationDetails}
We implement SemExplainer using PyTorch 1.10.0.
Mask optimization is performed on three datasets over 100 epochs.
To balance explainer performance with sparsity constraints, we set the sparsity coefficient to $\lambda$ = 0.1 by default. 
All quantitative experiments are conducted on a server equipped with two A100 GPUs. 
Each experiment is repeated four times, and we report the average results along with the standard error.
It is important to note that SemExplainer is an instance-level explainer. 
In other words, a single optimization run generates explanations for only one test instance. 
Consequently, during optimization, we don't distinguish between training and test sets in the three datasets. 
Instead, test samples are randomly selected and provided as input to SemExplainer.
For comparison experiments, directly using explanations generated by existing baseline methods to compute the SIS and SIN metrics is challenging. 
Simply treating the explanatory subgraphs generated by baseline methods as synergistic information and the irrelevant subgraphs as non-synergistic information would lead to unfair comparisons. 
This is because the explanatory subgraphs are designed to contain all synergistic information, while the irrelevant subgraphs consist solely of non-synergistic information. 
Consequently, the SIS and SIN metrics become ineffective. 
To address this issue, we randomly mask the explanations generated by the baseline method, splitting each explanatory subgraph into two parts. 
One part is treated as the synergistic subgraph and the other as the non-synergistic subgraph, based on which the SIS and SIN metrics are computed. 
This approach focuses on evaluating how well the explanatory subgraph captures synergistic effects without interference from irrelevant information.

\end{document}